# The dynamics of H-bonds of the hydration shells of ions, ATPase and NE-activated adenylyl cyclase on the coupling of energy and signal transduction

Alfred Bennun[1,2]

[1]Full Professor Graduate School, Rutgers University, USA
[2]CONICET, Argentina

## Abstract

Glycerol titration distinguished from free water the local hydration shell involved in ATPase transition from active to inactive, with cooperativity for water n=16. Rat brain cortex: NE-stimulated and its basal AC in the absence of free $Mg^{2+}$, allows a refractive state of AC with negative cooperativity for MgATP and $ATP^{4-}$. The erythrocyte-Hb system operates as a metabolic sensor to match glucose availability with the release of Hb-carried, $O_2$ and $Mg^{2+}$ at CSF. $[Mg(H_2O)_6](H_2O)_{12}^{2+}$ by chelating either a protein or $ATP^{4-}$ losses most of its hydration shell. The ion pump ATPase by forming $ADP^{3-}$ releases an incompletely hydrated $Mg^{2+}$, which could capture $H_2O$ from either $[Na.(H_2O)_6]^+$ or $[K.(H_2O)_6]^+$. Thus, sieve-sizing their hydration shells for fitting into $Na^+$-pump channels. An AC refractory period may participate in STM. $Mg^{2+}$ with cooperativity n=3.7 activates NE-AC. CREB-generated receptors coupled by $Mg^{2+}$ may modulate hydration shells-dependent oscillations for retrieval of LTM.

Key works: hydrated ATPase, refractory adenylate cyclase, hydration shells sizing, $Mg^{2+}$-coupling, Hb-$O_2$/$Mg^{2+}$-carrier, CREB stability, LTM retrieval.

It was undertaken the elucidation of the links between the hydration shell chemistry of proteins **[1]** in signal propagation and memory affirmation **[2] [3] [4]**. Experimental findings and theoretical ones were applied to characterize and model a mechanism of protein, ion-hydration coupling in molecular signal-electrogenic transduction.

The characterization of a norepinephrine (NE)-stimulated adenylyl cyclase (AC) [EC 4.6.1.1] of rat brain cortex and corpus striatum **[5]** indicated that cAMP could be involved in the function of memory pathways. This conjecture had experimental support from many laboratories and has been successfully generalized as behavior-cAMP linked models **[5] [6] [7] [8] [9] [10]**.

However, it also requires validation of physicochemical parameters, which allow encoding brain states, and their retrieval, at the molecular level **[7] [10] [11]**. An applicable mechanism has to restrict thermic equilibrium **[12] [2] [13] [14]**, which at the protein structure level, may result in a background random noise, equivalent to 0.7kcal/mol at 37°C.

The strength of H-bonds: O–H…: O (5.0 kcal/mol), N–H…: N (3.1 kcal/mol), N–H…:O (1.9 kcal/mol) HO–H…:$OH_3^+$ (4.3kcal/mol) **[15]** allows differentiating thermic-mediated oscillations **[16]** from hydration shell-mediated, oscillatory changes of molecular reactivity **[13]**. This suggests a major role for hydration-mediated conformational changes of proteins. These processes interact through the reordering of H-bonds for energy- and signal-transduction **[2] [17] [18]** in electrogenic membranes.

The lipid configuration of a cellular membrane allows proteins sidedness, which involves restricted water reactivity at hydrophobic local sites. The changes in quaternary structure, involves a dissipative potential, from a hydration shell as $H_2O$-donor to a subsequent acceptor shell of a protein. This process become eventually coupled to heat equilibrium with $H_2O$-cluster $(H_2O)_n$, H-bond dynamics and outflow through cerebrospinal fluid (CSF) renewal **[13] [14]**.

A spontaneous reactivity order with time restricted thermic equilibrium could be described by water exchanges between hydration shells. Chaotropic $[Mg(H_2O)_6](H_2O)_{12}^{2+}$ as a ligand with a protein loses most of its hydration sphere to form a bi-, tetra- or hexavalent coordinative chelate complex. Hence, when the latter breaks down $Mg^{2+}$, is released with a partially hydrated shell. The value of its ΔE or thermodynamic tendency **[15]** to complete its own shell, could be used to predict if and when $Mg^{2+}$ ions could subtract water from chaotropic $[Na(H_2O)_6]^+$, generating $[Na(H_2O)_3]^+$. The latter could subtract $H_2O$ from kosmotropic $[K(H_2O)_6]^+$ to form $[K(H_2O)_3]^+$. This dissipative ion-water structures conforms a thermodynamic **[13] [14]** chain which has as donor/receptor of last recourse becomes $(H_2O)_n$ **[19] [20] [21] [22]**.

Networks of interacting hydration shells, mediated by the structural changes of H-bonding in turnover, do appear capable to function as carriers of information-signaling and lead to evaluate its functional parameters. This may involve synchronizing the $Na^+/K^+$ ion pump **[23]**, with neurotransmitter dependent activation of adenylyl cyclase (NE-AC) **[5]**. The latter forms the cAMP required to act as promoters (CREB) **[9]**. These ones modulate neuronal DNA transcription **[6]**, to support a role in memory affirmation processes **[17]** involved in long-term memory (LTM) **[24] [3]**.

**Methods**

*Preparation of heat-activated ATPase*

The chloroplast membrane preparation of $CF_1$ (ATP synthase) containing negligible ATPase activity, was purified with DEAE-Sephadex A-50 and diluted in 20mM ATP in 10% v/v glycerol. A 2 mg/ml of this solution of $CF_1$-ATPase could be stored for months at -60°C. Defrosted and diluted samples of this preparation were used to prepare tritium-labeled acetyl-$CF_1$, which required $Mg^{2+}$ or $Ca^{2+}$ for its binding to $CF_1$-depleted chloroplast membranes. A heat-treatment of this preparation of latent-ATPase suppressed its $Mg^{2+}$-dependence for binding to an ATPase-depleted membrane, and its capacity to stimulate photophosphorylation [25]. To develop its $Ca^{2+}$-dependent ATPase activity, the $CF_1$ preparation was heated in a water bath at 65º for 3min, in a medium containing 20mM ATP, 1mM EDTA, 20mM $CaCl_2$, 0.2% bovine serum albumin, and 50mM Tris-Cl. Heating was stopped by a rapid cooling procedure, consisting of the addition of equal volume of ice-cold 0.1M Tris-Cl, pH=8. Thus, allowing to incubate at $37^oC$ the samples of $Ca^{2+}$-dependent ATPase, included in the mixtures reported under fig.1. The inorganic phosphate released was measured by the method of Chen [26].

*Preparation of EDTA-treated adenylyl cyclase*

The brain of rats killed by decapitation at $4^oC$ were sliced and separately collected to obtain cerebral cortex, hypothalamus, and corpus striatum slices in ice-cold 2mM-Tris/HCl buffer/10mM-NaCl/10mM-KCl/5μM-EDTA at final pH=7.4. The tissues were homogenized in a motor-driven glass homogenizer at 400rev/min for 2min. The homogenate was centrifuged at 2400rev/min for 20min and the supernatant fraction was discarded, the pellet was washed by resuspension with a vortex mixer. The washing procedure was repeated twice. This EDTA-treated adenylyl cyclase (AC) particulate preparations were stored at $-70^oC$ and used after thawing for 30min at $24^oC$ [5]. cAMP formed was measured by a modification of the Gilman method [27]. The assay reactants added to the reaction mixture: 1.25pmol of cyclic[$^3$H]AMP ($NH^{4+}$ salt containing 8000 c.p.m.) and 15 to 30 μg of binding protein prepared from fresh bovine muscle by the method of Miyamoto et.al. [28].

<div align="center">**Results**</div>

**Hydration shell modulation of the responsiveness of an ATPase active site**

A water requirement for the transition state from active to inactive was measured as a function of a glycerol saturation curve, on a previously purified heat-activated ATPase [8].

Fig.1.a) shows the obtained data in a Lineweaver-Burk plot, as a function of the ATPase activity in dependence of the molarity of $H_2O$. The mixed appearance of the plot fig.1.a), may be simply explained by the $H_2O$ saturation interactions at two very similar active sites of the protein molecule.

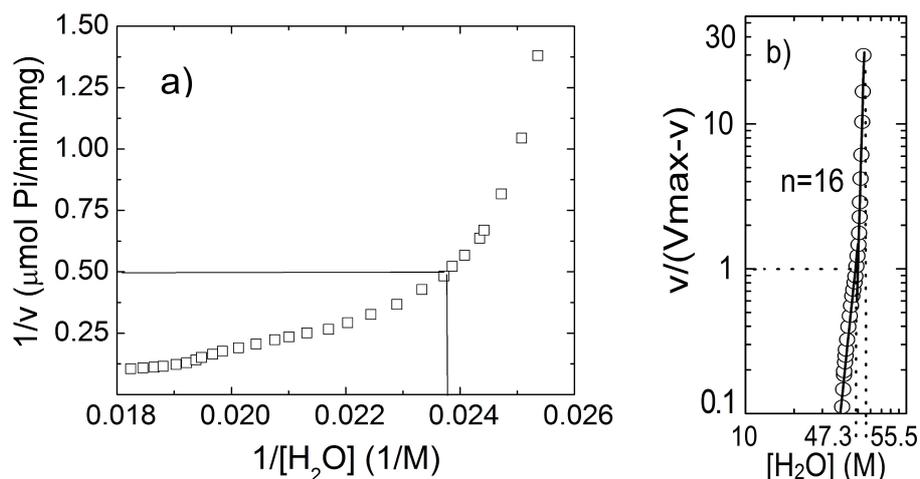

**Figure 1. a) Linewaever-Burk plot.** 3.7μg of heat-activated ATPase-CF$_1$ 50mM Tricine-NaOH, 8mM ATP-CaCl$_2$ pH=8 with glycerol additions, subtracting H$_2$O from the hydration shell of ATPase in a final 1.25ml were incubated at 37°C for 10min. Pi was measured and zero time blanks were done for each activity assayed. Since the plot resulted in a curve the initial ATPase activity: v=9.57 μmoles Pi/min/mg was used as Vmax app. **b) Hill Plot:** The concentration of pure water: [H$_2$O]=1000/18=55.5 M. Addition of 1.6 M glycerol decreases by 50% the optimal hydration state ATPase HS$_{50}$=49.5M H$_2$O. The initial activity was use as Vmax app. Applying Hill equation, $\log(v/(V_{max} - v)) = n \log[L] - \log[K_d]$, in which L=47.3/55.5, Kd(H$_2$O)=13.75, obtained n=16.

Applying the Hill plot, fig.1.b), it was possible to detect some superposition between two hydration tendencies, but the methodology was not sensitive enough to obtain separated hydration curves. Therefore, the Hill equation was used to determine the cooperative number of H$_2$O required, to optimize the local hydration shell(s) of the active site(s) of ATPase. As a coordinative ligand of ATPase, H$_2$O interacts with a cooperative value: n=16.

Divalent ions like Mg$^{2+}$, when forming chelate-complexes like MgATP, must loss most of their hydration shells. Divalent metals tend to form with oxygen containing ligands, complexes resistant to dissociation like MgATP K$_d$=5x10$^{-5}$M [29]. Furthermore, multivalent metals could form chelating complexes with electronegative atoms, which are closed positioned by the R-groups within a protein. The α-carbons (Cα) of His-groups and other R-groups allow rotation when a chelate complex breakdown and releases the divalent metal, within an incompletely hydrated shell.

Therefore, completing the hydration shell of the ion Mg$^{2+}$ may depend from a water donor like the hydration spheres of other ions or proteins. The lipid enclosure of proteins within a membrane, may favor that incomplete hydrated Mg$^{2+}$ reacts with proteins than with (H$_2$O)$_n$. Resolution-reconstitution studies [25] show that ATPase requires Mg$^{2+}$ and a lipid extract, to bind with the resolved ATPase-depleted membrane. Homology between (Na$^+$+K$^+$), K$^+$- and Ca$^{2+}$-ATPases [30] allows to infer that in these enzymes, H$_2$O by forming hydration shells, stabilize protein conformational changes.

The chain of events indicates that the association-dissociation thermodynamic ordering may also implicate $(H_2O)_n$. The process of H-bonds break down and reorganization may follow a decrease in free energy. The strength of H-bonding of $H_2O$ between N atoms in an R-group of hemoglobin (Hb) is larger than the H-bond between two water molecules within a cluster. Stoichiometry may involve the breaking of more than two H-bonds in the cluster.

The stoichiometry of 16 $H_2O$ per one molecule of ATPase, shown at fig.1.b, indicates that conformational change related to the number of H-bonds decreasing in the cluster, could become potentiated by the number of $H_2O$ exchanged. Endergonic release of hydration water from proteins may depend of coupling with an *in common* exergonic agent. Ion $Mg^{2+}$ binding allow a signaling network controlled by the dissociation state of $Mg^{2+}$. Turnover of the protein-metal complex for releasing the metal may depend on the number of $H_2O$, participating in each of the protein states. Turnover shift between metal complexes and free proteins became balance by the hydration state allowing an overall decrease in free energy, driven by the entropy increase within $(H_2O)_n$.

According to its hydrated state a released $Mg^{2+}$ would have the tendency to increment water coordination to complete its hydration shell. The strength order for such process, is shown simplified by not including all intermediate species, follows: $[Mg(H_2O)_2]^{2+}$, … $[Mg(H_2O)_6]^{2+}$, and the water saturated specie …$[Mg(H_2O)_6](H_2O)_{12}^{2+}$, ΔE: 70.9, … 24.5 and 13.1 kcal/mol respectively **[15]**. The corresponding charge effective $q$(Mg): +1.64,… +1.29 and +1.13, respectively **[15]**. Hence, $Mg^{2+}$ shows versatility in its tendency for completion of its hydration shells. It is assumed a central role for $Mg^{2+}$ as a driving force, for coupling dipolar charge mediated conformational changes of membrane proteins **[18]**.

**$Mg^{2+}$-dependent responsiveness of adenylyl cyclase to neurotransmitter**

Adrenergic nerve fibers make up the sympathetic nervous system which releases the neurotransmitter norepinephrine (NE) at the synapse, or junction. Alterations in the endogenous cyclic AMP could be caused by changes in metabolic conditions, for example, as the inhibition of phosphodiesterase-dependent breakdown of cyclic AMP. The isolation from rat's brain of adenylyl cyclase (AC) [EC 4.6.1.1] and characterization for its neurotransmitter responsiveness **[5]**, allowed evaluation of a possible role of the enzyme in cAMP-dependent memory processes. NE activation, when $Mg^{2+}$ is in excess of substrate, of an isolated AC from brain, was first reported for cortex and corpus striatum **[5] [31]**. The enzyme relates to a network of tissues **[32] [33]** supporting brain activity **[24]** by their *in common* response to the modification of free ionic $Mg^{2+}$, by changes in the concentration of chelating metabolites.

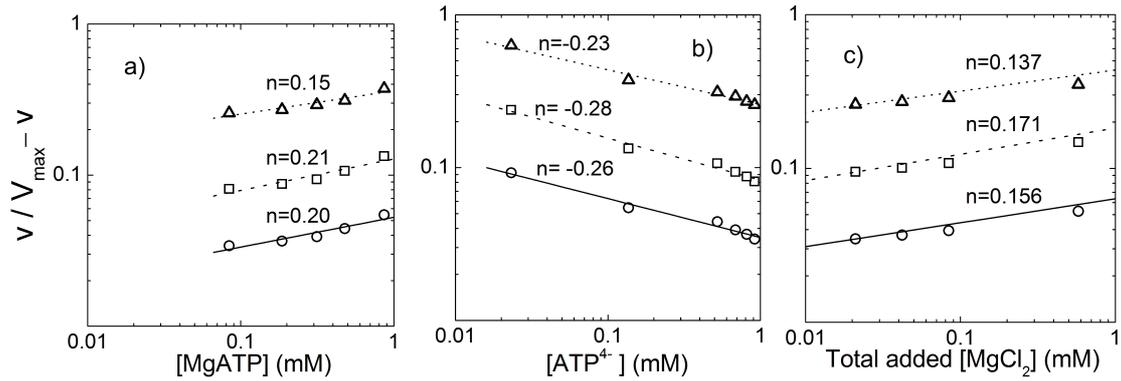

**Figure 2. Hill Plots of the separate effects of: (a) MgATP, (b) ATP$^{4-}$ and (c) Total added MgCl$_2$, on AC when the concentration of Mg$^{2+}$ does not exceed the one required to form substrate MgATP**. 100μg samples of EDTA-treated membrane preparations of rat's brain cortex were assayed for AC-activity at constant 1mM-ATP$^{4-}$ and increasing MgCl$_2$ without exceeding the one required for the formation of the substrate MgATP = Total added MgCl$_2$. Responsiveness to: □) Total MgCl$_2$ basal (4); o) + 0.1mM-norephinephrine (4); Δ) : +0.1mM-norephinephrine and +0.3mM-CaCl$_2$ (4). Multiple equilibrium equations and dissociation constants for MgATP were solved to obtain the separate concentrations of a) MgATP, b) ATP$^{4-}$ at the total added c) MgCl$_2$.

The Hill Plots in fig.2 shows that by solving multiple equilibrium equations was possible to separate the effect of MgATP and ATP$^{4-}$ on AC activity, when total added MgCl$_2$ concentration does not exceeds the required for forming its substrate MgATP. Characterization of the enzyme incubated with NE, and the one with this neurotransmitter, plus added Ca$^{2+}$ indicate negative cooperativity for either the substrate MgATP, fig.2a), or ATP$^{-4}$, fig.2.b). This response to total MgCl$_2$ at constant 1mM-ATP$^{4-}$ fig.2.c) reflects that the scarce concentration of the dissociated specie Mg$^{2+}$ from MgATP Kd=5×10$^{-5}$ M. The ion Mg$^{2+}$ concentration in the mixture is shown as insufficient for reaching the concentration required, for a refractive form of the enzyme structure to become converted into an active Mg$^{2+}$-enzyme-complex **[31]**.

This property may allow an Mg$^{2+}$-dependent cycle of AC responsiveness. This one may allow a refractive period lacking AC activity allowing dependence from a brain area stimulus, leading to increase the demand for O$_2$ and Mg$^{2+}$ at CSF. At the neuronal cytoplasmic level and AC refractive state, may favor coupling with an Mg$^{2+}$ donor like the Na$^+$/K$^+$-ATPase. The latter, catalyzes the breakdown of MgATP to forms MgADP, kd=10$^{-4}$ M, phosphate$^{2-}$ and incompletely hydrated free Mg$^{2+}$. Hence, it could be infer that the enzyme appears to be evolutionary adapted to participate in physiological refractive states. Thus, time-superpositioned events may be conflictive with LTM-processing, e.g., reflecting that vision-recording needs about 1/10 s.

Hence, a cycle of refractive versus active states of AC may lead to cytoplasmic modulation of cAMP level, which could become signal pulses. Fig.3 shows that a full stimulation effect on the enzyme by NE relates to the availability of free ion Mg$^{2+}$, saturating AC.

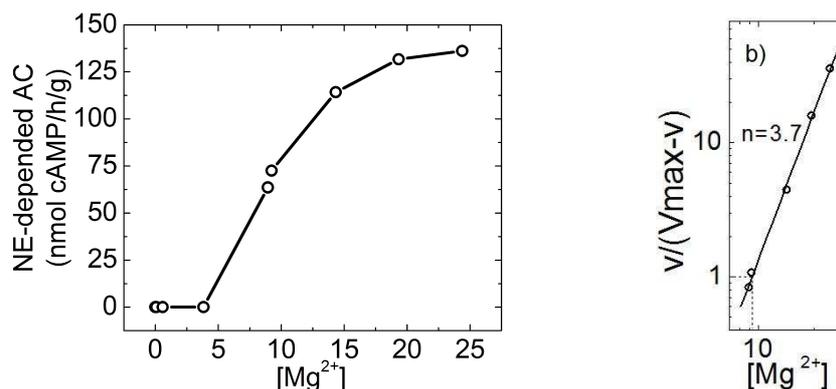

**Figure 3. a) Saturation curve of ion $Mg^{2+}$ in excess of constant $ATP^{4-}$ on adenylyl cyclase.** 100μg of protein from the EDTA-AC-membrane preparation in 40mM-tricina/Tris-buffer, pH= 7.4, 6.67mM-caffeine and constant 1mM ATP (Sigma: purified horse muscle containing 1.5%GTP) and addition of $MgCl_2$ for a saturation curve in 0.5ml final volume were incubated at 25°C for 1h in the presence of 0.1mM-norepinephrine (NE Vmax=214) and its absence (basal Vmax=76±7(4)). Plotted data o) the difference between: +0.1mM-norepinephrine and basal curves.

Fig.3.b) shows that the curve obtained from the difference between the NE-activated and the basal saturation curves, was plotted versus the free ionic concentration of $Mg^{2+}$, obtained by solving multiple equilibrium equations **[34] [35]**.

A basal sigmoidal cooperativity for $Mg^{2+}$ n=2.6 could result from the saturation requirement of the ion to increase the substrate concentration of MgATP and the decrease of $ATP^{4-}$ inhibitory effect, plus a requirement for the cyclizing reaction and/or activatory site **[31]**.

$Ca^{2+}$ has been shown for the transmembrane AC of brain to be inhibitory **[5]**. $Ca^{2+}$ has been postulated to function as activatory, for a different soluble enzyme, sAC, in other physiological systems **[36]**.

In presence of NE the brain AC shows cooperativity for $Mg^{2+}$ n=3.7, fig.3.b), indicating other participating $Mg^{2+}$ interacting sites in addition to these present for basal AC. The hormonal AC activity of fat cells shows a similar or higher $Mg^{2+}$ cooperativity values, for several activatory hormones **[32]**. Hence, an additional $Mg^{2+}$ binding site could be postulated to mediate the coupling of the NE receptor with AC and/or for the binding of AC to the electrogenic membrane. A divalent metal participation in the binding of membrane receptors includes GTP-binding regulatory proteins, ionotropic receptors, oxytocin receptor, etc.

**Hemoglobin efficiently transports $O_2$ and $Mg^{2+}$, which by 2,3-DPG mutual exclusion are released within CSF**

The Hb structure is well known and its cooperativity for $O_2$ has been widely documented **[37]**. Homotropic models will surge from the equilibrium of Hb oxygenation as the function of only $pO_2$ and $\Delta pH$. Physiologically, heterotropic interactions of Hb could result from the presence of the $O_2$-binding inhibitor 2,3-DPG and the divalent metals $Zn^{2+}$ and $Mg^{2+}$ in the erythrocyte **[37] [38] [39]**. The relationship between structure and function of Hb was examined to show evidence for Hb acting as a carrier of $Mg^{2+}$ **[40]**. Hence, allowing a homeostatic matching of $4O_2$ and $2Mg^{2+}$, to supply the requirements for aerobic glycolysis and for $Mg^{2+}$ availability in support of AC neurotransmitter responsiveness.

The denomination of histidines R-group as $Zn^{2+}$- as well as $Mg^{2+}$-fingers could be applied to their role, within Hb, in the course of its oxygenation vs. de-oxygenation processes, fig.3. The complexes $(O_2)_4Hb\text{-}(Zn^{2+})_2$ and $(O_2)_4Hb\text{-}(Mg^{2+})_2$ formed in the lungs, interact at lower $pO_2$ with 2,3-DPG to form 2,3-DPG-deoxyHb complexes, forcing the release of partially hydrated $Mg^{2+}$ and $O_2$ in CSF **[37] [40] [41] [42]**.

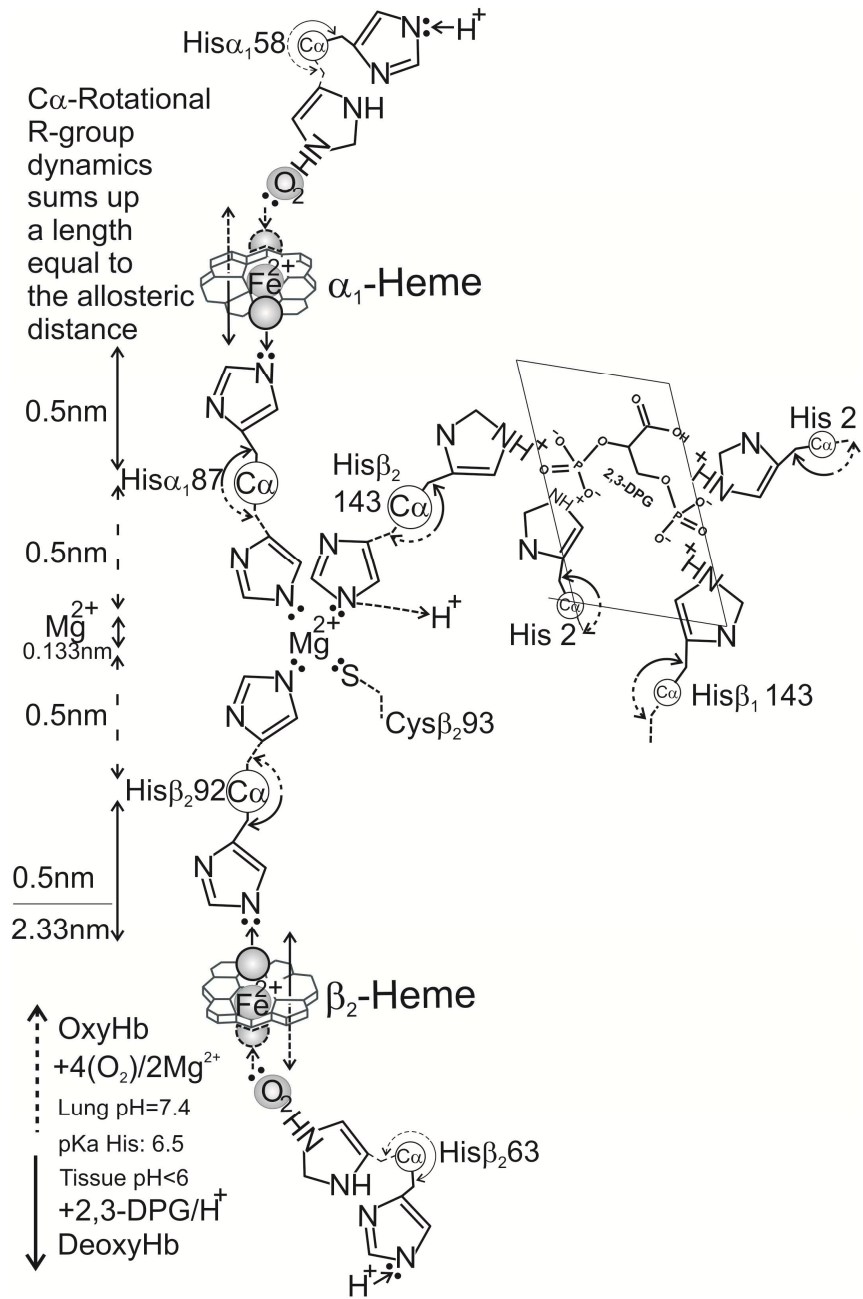

**Figure 4. Hb model illustrating that Cα–dynamics by rotating R-groups could reach the allosteric distance between a β₂-Heme and a α₁-Heme.** Only the interaction at the β₂α₁ interphase is illustrated. Hb crevices hold the two β-Hemes in between the groups Hisβ63 and Hisβ92, and the two α-Hemes in between His58 and His87. In the oxygenation sequence (in broken lines: ------) the $Fe^{2+}$ atoms which in the deoxyHb (◯) were within the internal face of Hemes move to the external one to react with $O_2$ (◌).

Fig.4 shows $O_2$ forming H-bonds with distal histidines: Hisβ₂63, Hisα₁58, Hisβ₁63 and Hisα₂58. One divalent metal, $Zn^{2+}$ or $Mg^{2+}$, fits at the interphase β₂α₁ and another divalent metal at β₁α₂.

Fig.4 shows that during oxygenation the proximal His β₂92 could no longer attract the Heme $Fe^{2+}$ because by Cα-rotation could participate jointly with Hisβ₂143 in the chelation of one $Mg^{2+}$ atom. Another divalent metal became chelated by $His\beta_1 92$ and $His\beta_1 143$.

Hence, in the deoxygenation events (in solid lines: ——) the proximal histidines reverse $C\alpha$-rotation reaching positions allowing to attract the $Fe^{2+}$ atoms.

Also, both $His\beta 143$ groups became released from the metal chelating sites and participate in the binding of 2,3-DPG.

Fig.4 illustrates that the participation of $Zn^{2+}$ and/or $Mg^{2+}$ in the dynamical integration of R-groups within the subunits interphase, allows to cover the allosteric distance of 2.33nm between Heme $\beta_2$ to Heme $\alpha_1$, and the similar one between Heme $\beta_1$ to Heme $\alpha_2$. Thus, allowing that the $Fe^{2+}$ became less tilted and positioned within the outer side of the Heme-plane for oxygenation

In peripheral tissues the deoxygenation events are promoted by a pH decrease, which allows an N atom of the amphoteric imidazole ring with pKa of 6.5 to bind protons, and react as an imino group of decreased affinity for divalent metals. However, at the CSF homeostatic pH=7.4 the increase in 2,3-DPG, attracts both $\beta_2$ and $\beta_1$ His143, fig.4, which by $C\alpha$-rotation became part of a single 2,3-DPG binding pocket. This one additionally conformed by the His 2 and Lys 82 of the two β chains. DeoxyHb became stabilized by salt links between $Asp\beta\ 94$ and $His\beta\ 146$ located in the same chain at both $\beta_1$ and $\beta_2$ chains. Additionally, salt links are formed between subunits as described and in the corresponding reverse ordering: $\alpha_1$ terminal-$COO^-$ …$NH_3^+$-$\alpha_2$ terminal, $\alpha_2$ 126 $Asp^-$ …$Arg^+$ $\alpha_1$ 141, plus $\beta_1$ terminal-$COO^-$ …$Lys^+$ 40 $\alpha_2$ **[37]**. Oxygenation breaks down these salt links of deoxyHb, the one in the β subunits between Asp94 and His146 became coordinate to the $Mg^{2+}$ ion **[37]**.

Hence, the ligands mass-action lead Hb to an equilibrium response for mutually inclusive interactions of increasing the divalent metals concentration and $pO_2$, allowing the simultaneous exclusion of 2,3-DPG and $H^+$ **[40]**.

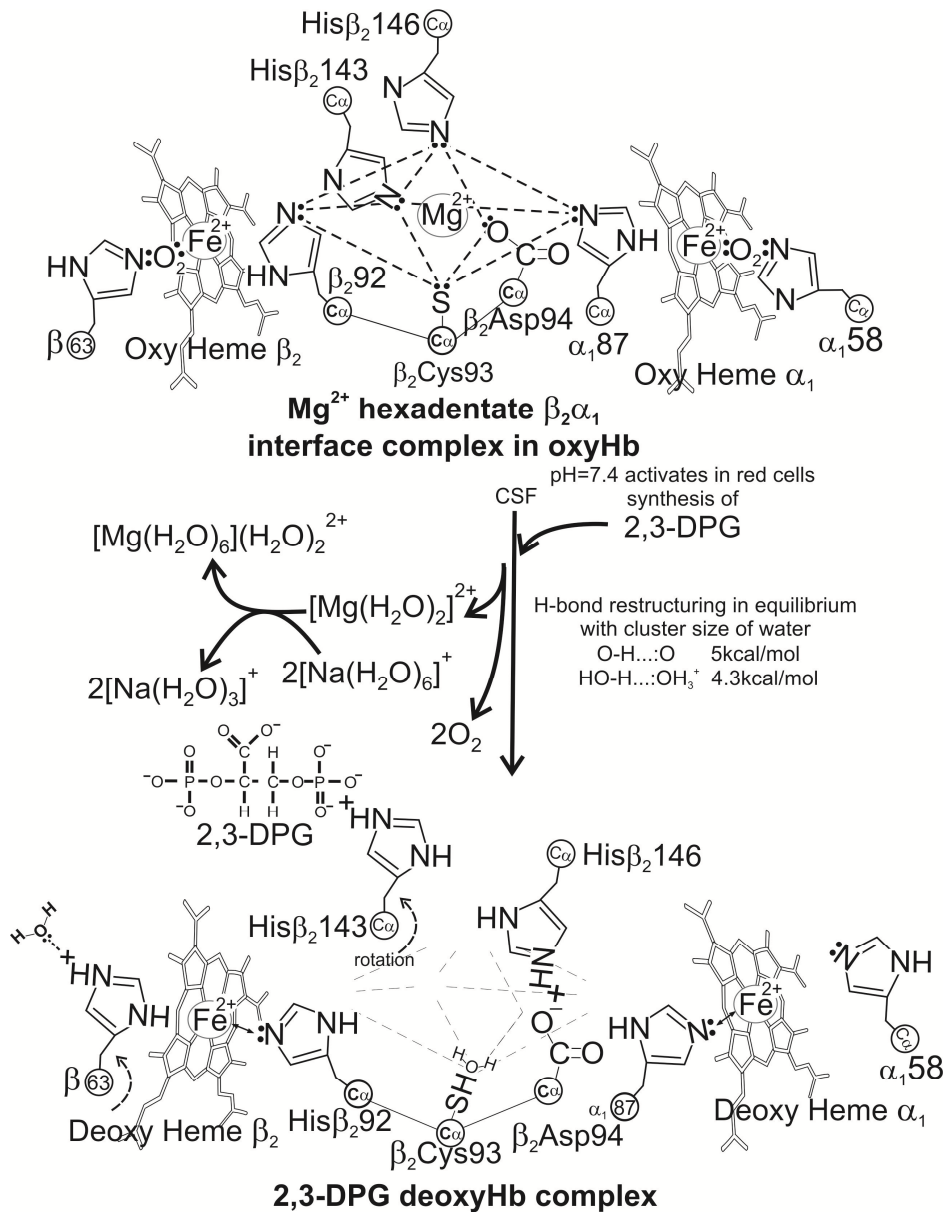

**Figure 5. At the CSF level the exergonic binding of 2,3-DPG to Hb is coupled with the endergonic release of $O_2$ and $Mg^{2+}$.** The proximal histidines, $His\beta 92$ and $His\alpha 87$, during binding of 2,3-DPG became released from the $Mg^{2+}$ binding site to attract Heme $Fe^{2+}$ promoting deoxygenation. The endergonic release of $Mg^{2+}$ became balance by the exergonic coordination of water molecules with the negative atoms of the R-groups. The N atom of a His R-group reacts with $H_2O$ for the exergonic formation of an H-bond: O-H$\cdots$N 6.9 kcal/mol, which could be compared with the endergonic breakdown of an H-bond in a cluster $(H_2O)_n$: O–H$\cdots$O 5.0 kcal/mol. Hence, the overall balance of H-bond formation and breakdown attenuates heat-exchanges, maintaining temperature homeostasis. The typical length of an H-bond in water is 0.197nm.

The amphoteric proximal His R-group by C$\alpha$-rotation leaves its position at $5^{th}$ coordination position of the $Fe^{2+}$ to enter into the $Mg^{2+}$-hexa-chelating structure shown in fig.5. As illustrated oxygenation promote the octahedral arrangement of Hb R-groups around $Mg^{2+}$, which allows the rupture of the salt links between Aspβ94 and His146 previously stabilizing deoxyHb.

A sensor role of the erythrocyte-Hb system responds to an increase in the level of glucose in blood, increasing its uptake by the red cell. The increase in the sugar phosphates intermediates of anaerobic glycolysis reduces the endogenous concentration of free divalent metals. This effect inhibits the $Mn^{2+}$-dependent peroxidase activity and decreases the HMP pathway generation of NADPH **[43]** and reduces the activity of the glyceraldehyde 3-phosphate dehydrogenase.

At the level of the CSF, the 2,3-DPG mutase of the Rapoport-Luebering glycolytic shunt has an activating optimum pH=7.4 and from 1,3-phosphoglycerate forms 2,3-DPG.

The latter, an Hb ligand with a $\Delta G = -5.7 Kcal/mol$ may bind first with $Hisβ_2143$ initiating a decrease in the chelating interactions of Hb with $Mg^{2+}$. As illustrated in fig.5 there is no need of a pH change to drive deoxygenation of Hb. This is done, by a progression of conformational changes, allowing at both Hb-interphases that the His2 and Lys82 complete the R-group structure holding in a single pocket 2,3-DPG to form the 2,3-DPG-deoxyHb complex. The binding of 2,3-DPG shows a mutually exclusive relationship with that of $O_2$ and $Mg^{2+}$ **[40]**.

Cooperativity may be reflected in the progression of the 2,3-DPG binding event inducing conformational changes. In this the tendency of $H_2O$ to form H-bonds with the negative atoms of one of the R-groups, facilitates similar effect on the other R-groups. This effect progressively reduces from hexa-, to tetra-, to bi-dentate, the metal-Hb-complex allowing a cooperative release of $O_2$ and $Mg^{2+}$.

**Kosmotropic versus chaotropic tendencies adjust the size of the hydration shells of $Na^+$ and $K^+$ ions for fitting into their gate-dynamical turnover at the $Na^+$-ion-pump**

Kosmotropes ions organize water into its hydration shell and have the tendency to subtract water from chaotrope ions like $[K(H_2O)_6]^+$. To the first group belong $[Na(H_2O)_6]^+$ with hexagonal geometry in the first water layer and $[Mg(H_2O)_6](12H_2O)^{2+}$ with octahedral geometry in the first and second layers.

Fig.6 illustrate that the differential strength between the tendencies of ions to complete their hydration shells allows directional reactivity. Moreover, the interactions of water depend of the different and variable stabilities for structuring H-bonds. Exclusion from hydrophobic regions of the protein-lipid electrogenic membrane and the proteins themselves allow one sided distribution of ligand water sites.

The dissipative states of water-protein structures **[14]** involve the break and reform of H-bonds, which occur within femtoseconds. If this were only a thermic event the molecule could have a non-local conformation change and a non-spatially directed vibrational state.

The nerve impulse allows $Na^+$ from the CSF to enter in the cytoplasm side of the membrane to manifest a positive charge. Charge rearrangement may allow that the $Mg^{2+}$ interacting with the membrane could be released as a less hydrated ion. Breaking a coordinative $Mg^{2+}$-membrane complex is endergonic and its forming is exergonic. This effect could couple charge exchanges with energy transduction as electrons displace within the electrogenic membrane.

If changes in hydration shells are inferred for ion translocation at the $Na^+$-pump **[23]** and at the ATPase, these two processes may cross-couple for partial loss of the hydric shells of $Na^+$ and $K^+$ to adapt their sizes **[44]** to corresponding gate (fig.6).

There could be partially hydrated species of $Mg^{2+}$, because to form a chelate the divalent metal has to loss most of its hydration shells. Therefore, when the MgATP, oxyHb-$Mg^{2+}$ and other $Mg^{2+}$-chelating metabolites complexes, break or dissociate could release $Mg^{2+}$, surrounded by an incompletely saturated hydric shell (fig.6).

Hence, an unsaturated hydric shell of the $Mg^{2+}$ ion is capable to subtract water from hexa-hydrated sodium ion, $[Na(H_2O)_6]^+$ to form $[Na(H_2O)_3]^+$, even if the more fully hydrated shell of $Mg^{2+}$, may loss water to a less hydrated specie of $Na^+$. The latter, chaotropic strength could capture $H_2O$ from the kosmotropic $[K(H_2O)_6]^+$ to generate tetrahydrated $[K(H_2O)_4]^+$ and trihydrated complexes $[K(H_2O)_3]^+$ (fig.6).

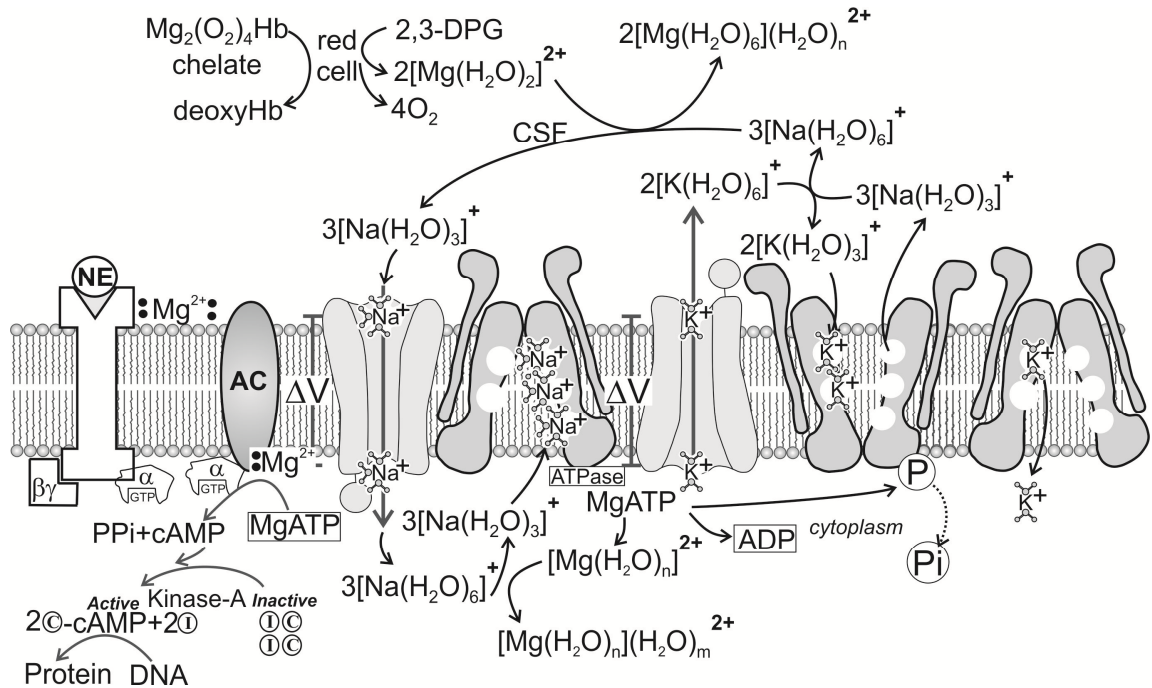

**Figure 6. Mg-driven Interaction of ion-hydration shells for the coupling of $Na^+/K^+$-ATPase and adenylyl cyclase...** The $Na^+$-in channel and the $K^+$-out channel open as a function of changing potential $\Delta V$. MgATP breakdown by the ion pump-ATPase is required for releasing $Mg^{2+}$ which when in excess of substrate became activatory of basal and NE stimulated AC.

The sizing of the hydration shell of $Na^+$ is widely accepted as required for the ion to penetrate into its channel at the $Na^+$-pump in order to reach the interior side of the membrane. The tendency to complete ion hydration shells became synchronized for nerve impulse. In the interior surface of the membrane $[(3H_2O).Na^+]$ should react with $[(6H_2O).K^+]$ and the generated $[(2H_2O).K^+]$ may be able to transit its channel **[44]**. The $Mg^{2+}$-driven hydric coupling for ionic translocation decreases the magnitude of $\Delta G<0$.

Dipolar molecular charge-cycles could allow to model the hydrated states of ions and protein dynamics of an ATPase for coupling with $Na^+/K^+$-pump turnover (fig.6) **[12] [23] [44]**. ATP's competition for an $Mg^{2+}$ chelated within the ATPase protein could be assumed to change the dipolar state of the ATPase. The ATPase activity may results in a coordinative phosphorylation, which may change electrostatic attractions into repulsions, within the subunits of the $K^+$ and $Na^+$ carriers **[12] [44] [45]**. The ATPase-dependent changes in the state of attraction between subunits of the carriers result in the appearance of dynamic structural modifications, which allow synchrony between $K^+$ uptake and $Na^+$ extrusion by the cell **[12]**. This also requires synchronization between availability of glucose and $Mg^{2+}$ **[46]**. If the glucose generated $ATP^{4-}$ exceeds the availability of $Mg^{2+}$ the effect could destabilize the $Mg^{2+}$-dependent binding of receptors to the membrane.

**Discussion**

The release of $Mg^{2+}$, from its chelating site at Hb, is an uphill thermodynamic event, which could be coupled to a decrease in free energy when 2,3-DPG binds to Hb. The overall process involves a decrease in free energy, because the $H_2O$ of the cluster could compete for the R-groups released in Hb when $Mg^{2+}$ and $O_2$ an release into the CSF-brain-system.

It is generally believe that synaptic plasticity implies structural changes that follow protein synthesis or in parallel. Molecular interaction affecting plasticity could involve formation of protein/enzymes complexes changing their heat-stability. It has been assume that a causal pathway for stress-pathologies may result from overexposure to epinephrine inactivating brain AC **[17]**.

Thermal-stability within limits may also have a participation in plasticity mechanisms, acting for selective memory processing. For example, euglena could synthetize *de novo* an induced-acid phosphatase of decreased thermal-stability, in response to phosphate deprivation/starvation, which is differentiable from the constitutive-enzyme **[47]**.

Thus, reversion mechanisms could add structural plasticity based in the differentiable perdurance of some protein structures.

This may become part of an overall recognition process in which the stimulus frequency for example, differentiate dim from brighter light by the number of the impulses conducted by the implicate retinal cells.

Proteins like ATPase, AC and $Na^+/K^+$-pump interacting with the associative-dissociative states of $Mg^{2+}$, are in a conformational turnover, which allows energy-transfer between molecules in the electrogenic membrane. Interatomic distances in the (donor protein)–H···(associate molecule) system: donor protein H-bonding distance ≈0.110 nm, whereas H···associate molecule distance is ≈0.160 to 0.200 nm. A non-equilibrium system surges if a protein, after being an optimal $H_2O$-donor by changing conformation, could become a poor $H_2O$-acceptor.

Restructuring of H-bonds, supporting the hydration shells of ions and proteins, could be regarded as dissipative states **[14] [48]** during turnover of the hydration-protein structures. The open system operation involves the uptake of glucose, which through aerobic glycolysis generates ATP as an enthalpy input, by its role as the chemical source of energy within the system. The output events became evident because the H-bonds thermodynamics maintains a no-equilibrium by incrementing entropy within the coupled $(H_2O)_n$.

A tight coupling, changing the hydration shells of protein molecules, could equilibrate with the clustering state of surrounding water, without significant disrupting the homeostatic brain temperature **[19]**. Hence, the capture of heat by the breaking of H-bonds results in smaller sized clusters, within the system. This is so, because the distance separating in a cluster $(H_2O)_n$, oxygen from oxygen, seeks to reach a local energy minimum, by changes in its charge distribution couple to its release of heat at lower temperature. The break and reform of H-bond in $(H_2O)_n$ occurs in $0.2 \times 10^{-12}$ s, but to reach an stationary state could take additional time **[25] [26]**.

The thermic equilibrium of hydration shells-$(H_2O)_n$ at CSF became retarded because the rate of CSF outflow from the brain system, allows its removal at the rate of 3.7 volumes of 160cm$^3$ of CSF renewal/day. The organismal open system allows outflow of water **[14]**. A retarded thermic equilibrium at the lower temperature, outside the body, allows reforming cluster sizes, because the dipolar electrostatic attraction between $H_2O$ molecules is a spontaneous process **[21] [22]**.

The turnover of the structural hydration of proteins conform a series of less and less reversible events. Vectorial kinetics becomes a manifestation of the no-equilibrium conditions maintained by the open system **[13] [14]**.

## Conclusions

Hb is a carrier discharging at CSF, not only $O_2$ but also $Mg^{2+}$, because both increase affinity for each other during Hb-oxygenation. The modeling of this process illustrates the Hb dynamics and shows that a protein could modify its internal R-groups reactivity, to allow intracoupling between activatory and inhibitory sites.

Deoxygenation of Hb releasing $4O_2$ and $2Mg^{2+}$ in CSF, (fig.4 and fig.5), may overcome the NA-AC refractory/latency condition (fig.2) by allowing $Mg^{2+}$-dependent coupling of AC with receptors to respond to neurotransmitter (fig.3).
A procedure differentiating ligand water from that in the surroundings of proteins is sought elsewhere with very expensive equipment, which indicates the advantage of the one described to obtain the results reported in fig.1.

At ATPase level, the glycerol methodology allowed to evaluate the number of water molecules, required to influence protein conformational changes. Hence, the finding illustrate that a turnover between states of hydration shells could result in molecular conformational oscillations. These ones are distinct and differentiable from temperature originated changes in rotation, translation and vibrational kinetics. The latter could be though as directionless and not leading to predictable or reproducible oscillations. Hence, it could not be expected that heat-randomness, could support vectorial kinetics for a specific signal-transmissions at the neuronal membrane.

It is proposed that H-bond kinetics allows molecular transitions to signal for information because allows the differentiation of vectorial processes from non-vectorial thermal tendencies to equilibrium **[13]**.

This conceptualization of a water mediated flow of energy, driving the non-equilibrium thermodynamics of electrogenic membranes, has been applied to schematize a consistent physicochemical modeling of neuronal activity.

This one allows to integrate the interaction of the hydric shells of enzymes and receptors, through a $Mg^{2+}$-mediated sizing of ions for their translocation within the $Na^+$-ion pump. Hydration shells-dependent changes modify the dipolar state of protein molecules within the membrane for charge-conductance. Thus, this modeling predicts that hydration-transfer processes could retrieve LTM information **[46] [4]**, to modulate the nerve impulse.

The modal model of short-term memory (STM), relates its capacity to about 7 elements per a period measure in seconds for preceding LTM **[49]**. STM implies subsequent neuronal changes to become stabilized by memory affirmation processes **[17]**. The latter, supports a behavior-cAMP linked model for LTM **[3] [4] [6] [7] [8] [9] [10] [11]**. Some researchers support a unitary memory hypothesis **[50]**. At the molecular level NA-AC negative cooperativity, could reflect the enzyme capability to receive signals preceding its activation (fig.2) and for activation (fig.3). Thus, the lack of free ion $Mg^{2+}$ may install a refractory period, which permits a variation of the rate of nerve impulse.

The binding of the newly synthetized proteins to the membrane **[25]** allows that by, stimulation through ion and/or charge displacements, the inserted proteins respond with changes in their hydric and dipolar states **[13]**. This produces conformational changes involved in the configuration of the H-bonds of receptors inserted in the neuronal membrane. These ones generate discrete oscillatory states, which as excited states of the electronic-lattices of inserted molecules within the membrane, may dissipate in about 2000ns. Quantum mechanics evaluates by their frequency distinct oscillatory states, to which assigns a wave function. These could be conserved for even longer periods if in resonance within a neuronal network.

**References**

[1] Wei, Y.Z., Kumbharkhane, A.C., Sadeghi, M., Sage, J.T., Tian, W.D., Champion, P.M., Sridhar, S. & McDonald, M.J.. Protein Hydration Investigations with High-Frequency Dielectric Spectroscopy. *J. Phys. Chem.* **98** (26), 6644–6651 (1994)

[2] Nestler, E.J. & Duman, R.S. Neurotransmitter and Signal Transduction. Section 1, Neuropsychopharmacology: The Fifth Generation of Progress, 5th edition. Editorial: Lippincott Williams & Wilkins (LWW) (2002)

[3]   Barco, A., Bailey, C. & Kandel, E. Common molecular mechanisms in explicit and implicit memory. *J. Neurochem.*, 97(6), 1520-33 (2006)

[4]   Bean, B.P. The action potential in mammalian central neurons. *Nature reviews. Neuroscience* **8**, 451-65 (2007)

[5]   Brydon-Golz, S., Ohanian, H. & Bennun, A. Effects of noradrenaline on the activation and the stability of brain adenylate cyclase. *Biochem. J.* **166**, 473-483 (1977)

[6]   Silva, A.J., Kogan, J. H., Frankland, P.W. & Kida, S. CREB and memory. *Annual Review of Neuroscience* **21**, 127-148 (1998)

[7]   Davis, H. P. & Squire, L. R. Protein synthesis and memory: a review. *Psychol. Bull.* **96**, 518–559 (1984)

[8]   Mayr, B. & Montminy, M. Transcriptional regulation by the phosphorylation-dependent factor CREB. *Nat. Rev. Mol. Cell Biol.* **2(8)**, 599-609 (2001)

[9]   Kandel, E.R. The molecular biology of memory: cAMP, PKA, CRE, CREB-1, CREB-2, and CPEB. *Mol Brain* **5**(1), 14 (2012)

[10]  Kandel, E. R. The molecular biology of memory storage: a dialogue between genes and synapses. *Science* **294**, 1030–1038 (2001)

[11]  Goelet, P., Castellucci, V. F., Schacher, S. & Kandel, E. R. The long and the short of long-term memory a molecular framework. *Nature* **322**, 419–422 (1986)

[12]  Bennun, A. The unitary hypothesis on the coupling of energy transduction and its relevance to the modeling of mechanisms. *Ann N Y Acad Sci.* **227**:116-45 (1974)

[13]  Bennun, A. Hypothesis for coupling energy transduction with ATP synthesis or ATP hydrolysis. *Nature (New Biology)* **233**, 5-8 (1971)

[14]  Prigogine, I., Lefever, R., Goldbeter, A. & Herschkowitz-Kaufman. Symmetry breaking instabilities in biological systems. *Nature* **223**, 913-6 (1969)

[15]  Pavlov, M., Siegbahn, P. & Sandstron, M. Hydration of Beryllium, Magnesium, calcium, and zinc ions using density functional. *J. Phys. Chem. A.* **102**, 219-228 (1998)

[16]  Rutishauser U., Ross I.B., Mamelak A.N., Schuman E.M., "Human memory strength is predicted by theta-frequency phase-locking of single neurons", *Nature* **464** (7290): 903–907 (2010)

[17]  Bennun, A. Characterization of the norepinephrine-activation of adenylate cyclase suggests a role in memory affirmation pathways. Overexposure to epinephrine inactivates adenylate cyclase, a causal pathway for stress-pathologies. *Biosystems* **100**(2), 87-93 (2010)

[18]  Bennun, A. Hypothesis on the role of liganded states of proteins in energy transducing systems. *Biosystems* **7**(2), 230-44 (1975)

[19]  Maheshwary, S., Patel, N., Sathyamurthy, N., Kulkarni, A. D. & Gadre, S. R. Structure and Stability of Water Clusters $(H_2O)_n$, n = 8-20: An *Ab Initio* Investigation. *J. Phys. Chem.* **105** (46): 10525 (2001)

[20]  Fanourgakis, G. S., Aprà, E., de Jong, W.A. & Xantheas, S.S. High-level ab initio calculations for the four low-lying families of minima of $(H_2O)_{20}$. II. Spectroscopic signatures of the dodecahedron, fused cubes, face-sharing pentagonal prisms, and edge-sharing pentagonal prisms hydrogen bonding networks. *J. Chem. Phys.* **122** (13), 134304 (2005)

[21]  Mir, M.H. & Vittal, J.J. Phase Transition Accompanied by Transformation of an Elusive Discrete Cyclic Water Heptamer to a Bicyclic $(H2O)7$ Cluster. *Angew. Chem. Int. Ed.* **46** (31), 5925–5928 (2007)

[22]  Smith, J.D., Cappa, C.D., Wilson, K.R., Cohen, R.C., Geissler, P.L. & Saykally, R.J. Unified description of temperature-dependent hydrogen-bond rearrangements in liquid water. *Proc. Natl. Acad. Sci. USA* **102** (40): 14171–14174 (2005)

[23] Skou, J.C. The influence of some cations on an adenosine triphosphatase from peripheral nerves. *Biochim. Biophys. Acta* **23**, 394 (1957)

[24] Bennun, A. Molecular Mechanisms Integrating Adenylyl Cyclase Responsiveness to Metabolic Control on Long-Term Emotional Memory and Associated Disorders (pp. 1-44). Chapter 1 in Long-Term Memory: Mechanisms, Types and Disorders Editors: Arseni K. Alexandrov and Lazar M. Fedoseev (2012)

[25] Bennun, A. & Racker, E. Partial resolution of the enzymes catalysing photophosphorylation IV. Interaction of coupling factor I from chloroplast with components of the chloroplast membrane. *J. Biol. Chem.* **244**, 1325-1331 (1969)

[26] Chen, P.S., Toribara, T.Y. & Warner H. Microdetermination of phosphorus. *Anal. Chem.* **28**, 1756-1758 (1956)

[27] Gilman, AG. A protein binding assay for adenosine 3':5'-cyclic monophosphate. *Proc Natl Acad Sci U S A* **67**(1), 305-312 (1970)

[28] Miyamoto, E., Kuo, J. F. & Greengard, P. J. Cyclic nucleotide-dependent protein kinases. 3. Purification and properties of adenosine 3',5'-monophosphate-dependent protein kinase from bovine brain. Biol. Chem. **244,** 6395–6402 (1969)

[29] Gupta, R.K., Gupta, P., Yushok, W.D. & Rose, Z.B. Measurement of the dissociation constant of MgATP at physiological nucleotide levels by a combination of $^{31}$P NMR and optical absorbance spectroscopy. *Biochemical and Biophysical Research Communications* **117**, 210–216 (1983)

[30] Serrano, R, Kielland-Brandt, MC & Fink, GR. Yeast plasma membrane ATPase is essential for growth and has homology with (Na+ + K+), K+- and Ca2+-ATPases. *Nature* **319** (6055) (1986)

[31] Ohanian, H., Borhanian, K., De Farias, S. & Bennun A. A model for the regulation of brain adenylate cyclase by ionic equilibria. *J. Bioenergetics and Biomembranes* **13**, 5-6 (1981)

[32] Harris, R. H., Cruz, R. & Bennun, A. The effect of hormones on metal and metal-ATP interactions with fat cell adenylate cyclase. *BioSystems* **11**, 29-46 (1979)

[33] Vicario, P.P., Saperstein, R., Bennun, A. Role of divalent metals in the kinetic mechanism of insulin receptor tyrosine kinase. *Arch. Biochem. Biophys.* **261**(2), 336-45 (1988)

[34] Taqui-Khan, M.M. and Martell, A.E. J. Phys. Chem. 66, 10-15 (1962)

[35] Taqui-Khan, M.M. and Martell, A.E.,J. Am. Chem. 88, 668-671 (1966)

[36] Tresguerres, M., Levin, L.R. & Buck, J. Intracellular cAMP signaling by soluble adenylyl cyclase. *Kidney Int*. **79**(12), 1277–1288 (2011)

[37] Fermi, G. & Perutz, M.F. In atlas of molecular structures in Biology, 2 Heamoglobin & Myoglobin. Phillips, D.C. and Richards F.M. (Eds.) Clarendon Press, Oxford. 4-5 (1982)

[38] Petersen, A, Kristensen, S.R., Jacobsen, J.P. & Hørder, M. $^{31}$P-NMR measurements of ATP, ADP, 2,3-diphosphoglycerate and Mg2+ in human erythrocytes. *Biochim Biophys Acta* **1035**(2), 169-74 (1990)

[39] Kovalevsky, A.Y., Chatake, T., Shibayama N., Park, S.-Y., Ishikawa, T., Mustyakimov, M., Fisher, S. Z., Langan, P. & Morimoto, Y. Preliminary time-of-flight neutron diffraction study of human deoxyhemoglobin. *Acta Cryst.* F**64**, 270-273 (2008)

[40] Bennun, A., Seidler, N. & De Bari, V.A. A model for the regulation of haemoglobin affinity for oxygen. *Biochemical Society Transactions* **13**, 364-366 (1985)

[41] Ronda, L., Stefano Bruno, Stefania Abbruzzetti, Cristiano Viappiani, Stefano Bettati, Ligand reactivity and allosteric regulation of hemoglobin-based oxygen carriers. *Biochimica et Biophysica Acta (BBA) - Proteins & Proteomics* **1784**, Issue 10, 1365–1377 (2008)

[42] Resina A, Brettoni M, Gatteschi L, Galvan P, Orsi F, Rubenni MG  Changes in the concentrations of plasma and erythrocyte magnesium and of 2,3-diphosphoglycerate during a period of aerobic training. Eur J Appl Physiol Occup Physiol. 1994;68(5):390-4.

[43] Bennun, A., Needle, M. & DeBari. V. Stimulation of the hexone monophosphate pathway in the human erythrocyte by $Mn^{2+}$: evidence for a $Mn^{2+}$-Dependent NADPH peroxidase activity. *Biochemical medicine* **33**, 17-21 (1985)

[44] Gadsby, D.C., Bezanilla, F., Rakowski, R.F., De Weer, P. & Holmgren M. The dynamic relationships between the three events that release individual $Na^+$ ions from the $Na^+/K^+$-ATPase. *Nature Cell Biology* **14**, 416–423 (2012)

[45] Garraham, P.J. and Glynn, I.M. Driving the sodium pump backwards to form adenosine triphosphate. *Nature* **211**, 1414-415 (1966)

[46] Guinzel, D. & Schlue, W.R. Sodium-magnesium antiport in Retzius neurones of the leech Hirudo medicinalis. *Journal of Physiology* **491**(3), 595-608 (1996)

[47] Bennun, A. & Blum, J.J. Properties of the induced acid phosphatase and of the constitutive acid phosphatase of euglena. *Biochim.Biophys.Acta* **128**, 106-123 (1966)

[48] Bennun, A. Primordial open-system thermodynamics and the origin of a biophysics selection principle. Open Journal of Biophysics, 2012, 2, ***-***. In press.

[49] Atkinson, R.C. & Shiffrin, R.M. Chapter: Human memory: A proposed system and its control processes". In Spence, K.W.; Spence, J.T.. The psychology of learning and motivation (Volume 2). *New York: Academic Press.* 89–195 (1968)

[50] Brown, G. D. A., Neath, I., & Chater, N. A ratio model of scale-invariant memory and identification. *Psychological Review*, 114, 539–576 (2007)